\documentclass[runningheads,a4paper]{llncs}
\usepackage{amssymb}
\usepackage{graphicx}
\usepackage{comment}
\usepackage{amssymb}
\usepackage{amsmath}
\usepackage{paralist}
\usepackage{stmaryrd}
\usepackage{multirow}
\usepackage{helvet}
\usepackage{courier}
\usepackage[title]{appendix}
\usepackage{graphicx}
\usepackage{algpseudocode}
\usepackage{graphics}
\usepackage{epsfig}
\usepackage{algorithm}
\usepackage{hyperref}
\usepackage{subfigure}
\usepackage{ifsym}
\usepackage{graphicx}

\usepackage{epstopdf}
\usepackage{xcolor}
\usepackage{tikz}
\usetikzlibrary{arrows}
\usetikzlibrary{decorations.markings}
\usetikzlibrary{shapes}
\usetikzlibrary{plotmarks}

\usepackage{algorithm,algorithmicx,algpseudocode}
\renewcommand{\algorithmiccomment}[1]{\bgroup\hfill\scriptsize//~#1\egroup}
\usepackage{multirow}
\usepackage{array,ragged2e}

\usepackage{color}
\usepackage{hhline}

\usepackage{pgfplots}

\usepackage{url}

\usepackage{tikz}
\usetikzlibrary{arrows,positioning,automata,decorations,fit,backgrounds,calc,shapes}
\usepgflibrary{shapes.geometric} 
\usetikzlibrary{shapes.geometric} 
\usepgflibrary{decorations.pathmorphing} 
\usetikzlibrary{decorations.pathmorphing} 

\usetikzlibrary{matrix}
\newcommand{\ANDAND}{\mathbin{\mathrm{AND}}}
\newcommand{\OPT}{\mathbin{\mathrm{OPT}}}
\newcommand{\bound}{\mathbin{\mathrm{bound}}}
\newcommand{\UNION}{\mathbin{\mathrm{UNION}}}

\newcommand{\FILTER}{\mathbin{\mathrm{FILTER}}}
\newcommand{\SELECT}{\mathop{\mathrm{SELECT}}}

\newcommand{\var}{\mathit{var}}

\newcommand{\true}{\mathit{true}}
\newcommand{\false}{\mathit{false}}
\newcommand{\error}{\mathit{error}}

\newcommand{\semm}[2]{\llbracket #1 \rrbracket_{#2}}

\newcommand{\dom}[1]{\mathrm{dom}(#1)}

\begin{document}

\mainmatter 

\title{PIWD: A Plugin-based Framework for Well-Designed SPARQL}
\titlerunning{PIWD: A Plugin-based Framework for Well-Designed SPARQL}
\authorrunning{X. Zhang et al.}
\author{Xiaowang Zhang\inst{1,3,4} \and Zhenyu Song\inst{1,3} \and Zhiyong Feng\inst{2,3} \and Xin Wang\inst{1,3}}
\institute{School of Computer Science and Technology, Tianjin University,
Tianjin, China
\and
School of Computer Software, Tianjin University, Tianjin 300350, China
\and
Tianjin Key Laboratory of Cognitive Computing and Application,
Tianjin, China
\and
Key Laboratory of Computer Network and Information Integration\\ (Southeast University), Ministry of Education, Nanjing 211189, China
}

\maketitle

\begin{abstract}
In the real world datasets (e.g.,DBpedia query log), queries built on well-designed patterns containing only AND and OPT operators (for short, WDAO-patterns) account for a large proportion among all SPARQL queries. In this paper, we present a plugin-based framework for all SELECT queries built on WDAO-patterns, named PIWD. The framework is based on a parse tree called \emph{well-designed AND-OPT tree} (for short, WDAO-tree) whose leaves are basic graph patterns (BGP) and inner nodes are the OPT operators. We prove that for any WDAO-pattern, its parse tree can be equivalently transformed into a WDAO-tree. Based on the proposed framework, we can employ any query engine to evaluate BGP for evaluating queries built on WDAO-patterns in a convenient way. Theoretically, we can reduce the query evaluation of WDAO-patterns to subgraph homomorphism as well as BGP since the query evaluation of BGP is equivalent to subgraph homomorphism. Finally, our preliminary experiments on gStore and RDF-3X show that PIWD can answer all queries built on WDAO-patterns effectively and efficiently.
\end{abstract}

\section{Introduction}\label{sec:introduction}
Resource Description Framework (RDF) \cite{Swick1998Resource} is the standard data model in the semantic web. RDF describes the relationship of entities or resources using directed labelling graph. RDF has a broad range of applications in the semantic web, social network, bio-informatics, geographical data, etc. \cite{Arenas2011Querying}. The standard query language for RDF graphs is SPARQL \cite{Prud2006SPARQL}. Though SPARQL is powerful to express  queries over RDF graphs \cite{Angles2008The}, generally, the query evaluation of the full SPARQL is PSPACE-complete \cite{P2009Semantics}.

Currently, there are some popular query engines for supporting the full SPARQL such as Jena \cite{Carroll2004Jena} and Sesame \cite{Broekstra2002Sesame}. However, they become not highly efficient when they handle some large RDF datasets\cite{Zou2011gStore,Zou2014gStore}.
Currently, gStore\cite{Zou2011gStore,Zou2014gStore} and RDF-3X\cite{Neumann2010The} can highly efficiently query large datasets. But gStore and RDF-3X merely provide querying services of BGP. Therefore, it is very necessary to develop a query engine with supporting more expressive queries for large datasets.

Since the OPT operator is the least conventional operator among SPARQL operators \cite{Zhang2014ipl}, it is interesting to investigate those patterns extending BGP with the OPT operator. Let us take a look at the following example.

An RDF example in \autoref{blogger.rdf} describes the entities of bloggers and blogs. The relationship between a blogger and a blog is revealed in the property of \emph{foaf:maker}. Both blogger and blog have some properties to describe themselves. Triples can be modeled as a directed graph substantially.
\begin{table}[h]
\centering
\caption{bloggers.rdf \label{blogger.rdf}}
\begin{tabular}{l|l|l}
\hline
Subject &Predict &Object \\ \hline
id1 &foaf:name &Jon Foobar \\ \hline
id1 &rdf:type &foaf:Agent \\ \hline
id1 &foaf:weblog &foobar.xx/blog \\ \hline
foobar.xx/blog &dc:title &title \\ \hline
foobar.xx/blog &rdfs:seeAlso &foobar.xx/blog.rdf \\ \hline
foobar.xx/blog.rdf &foaf:maker &id1 \\ \hline
foobar.xx/blog.rdf &rdf:type &rss:channel \\ \hline
\end{tabular}
\end{table}

\begin{example}\label{ex:introduction}
Consider the RDF dataset $G$ storing information in \autoref{blogger.rdf}.
Given a BGP $Q=((?x, \textit{foaf:maker}, ?y) \ \ANDAND \ (?z, \textit{foaf:name}, ?u))$, its evaluation over $G$ is as follows:
\begin{center}
$\semm {Q}{G} =$\;\;\;\;\;\;\;\begin{tabular}{|c|c|c|c|}
\hline
?x &?y &?z &?u \\ \hline
foobar.xx/blog.rdf &id1 &id1 &Jon Foobar \\ \hline
\end{tabular}
\end{center}

Consider a new pattern $Q_1$ obtained from $Q$ by adding the OPT operator in the following way:
\\$Q_1=(((?x, \textit{foaf:maker}, ?y) \ \OPT \ (?y, \textit{rdf:type}, ?v)) \ \ANDAND \ (?z, \textit{foaf:name}, ?u))$, the evaluation of $Q_1$ over $G$ is as follows:
\begin{center}
$\semm {Q_1}{G} =$\;\;\;\;\;\;\;\begin{tabular}{|c|c|c|c|c|}
\hline
?x &?y &?v &?z &?u \\ \hline
foobar.xx/blog.rdf &id1 &foaf:Agent &id1 &Jon Foobar \\ \hline
\end{tabular}
\end{center}

Consider another pattern $Q_2=(((?x, \textit{foaf:maker}, ?y) \ \OPT \ (?y, \textit{rdf:type}, ?z)) \ \ANDAND \ (?z, \textit{foaf:name}, ?u))$, the evaluation of $Q_2$ over $G$ is the empty set, i.e., $\semm {Q_2}{G} = \emptyset$.
\end{example}

In the above example, $Q_1$ is a well-designed pattern while $Q_2$ is not a well-designed pattern \cite{P2009Semantics}.

In fact, we investigate that queries built on well-designed patterns are very popular in a real world. For example, in LSQ\cite{saleem2015lsq}, a Linked Dataset describing SPARQL queries extracted from the logs of four prominent public SPARQL endpoints containing more than one million available queries shown in \autoref{lsq}, queries built on well-designed patterns are over 70\% \cite{HAN2016On,Song2016Efficient}.
\begin{table}[h]
\centering
\caption{SPARQL logs source in LSQ \label{lsq}}
\begin{tabular}{l|l|r}
\hline
Dataset &Date &Triple Number \\ \hline
DBpedia &30/04/2010 to 20/07/2010 &232,000,000 \\ \hline
Linked Geo Data (LGD) &24/11/2010 to 06/07/2011 &1,000,000,000 \\ \hline
Semantic Web Dog Food (SWDF) &16/05/2014 to 12/11/2014 &300,000 \\ \hline
British Museum (BM) &08/11/2014 to 01/12/2014 &1,400,000 \\ \hline
\end{tabular}
\end{table}

Furthermore, queries with well-designed AND-OPT patterns (for short, WDAO-patterns) are over 99\% among all queries with well-designed patterns in LSQ \cite{HAN2016On,Song2016Efficient}. In short, the fragment of WDAO-patterns is a natural extension of BGP in our real world. Therefore, we mainly discuss WDAO-patterns in this paper.

In this paper, we present a plugin-based framework for all SELECT queries built on WDAO-patterns, named PIWD. Within this framework, we can employ any query engine evaluating BGP for evaluating queries built on WDAO-patterns in a convenient way. The main contributions of this paper can be summarized as follows:
\begin{compactitem}
\item We present a parse tree named \emph{well-designed AND-OPT tree} (for short, WDAO-tree), whose leaves are BGP and all inner nodes are the OPT operator and then prove that for any WDAO-pattern, it can be translated into a WDAO-tree. 
\item We propose a plugin-based framework named \emph{PIWD} for query evaluation of queries built on WDAO-patterns based on WDAO-tree. Within this framework, a query could be evaluated in the following three steps: (1) translating that query into a WDAO tree $T$; (2) evaluating all leaves of $T$ via query engines of BGP; and (3) joining all solutions of children to obtain solutions of their parent up to the root. 
\item We implement the proposed framework PIWD by employing gStore and RDF-3X and evaluate the experiments on LUBM.
\end{compactitem}

The rest of this paper is organized as follows: Section \ref{sec:pre}
briefly introduces the SPARQL, conception of well-designed patterns and OPT normal form.
Section \ref{sec:wdao} defines the well-designed and-opt tree to capture WDAO-patterns. Section
\ref{sec:piwd} presents PIWD and Section \ref{sec:experiment} evaluates
experimental results. Section \ref{sec:related} summarizes our related works. Finally, Section \ref{sec:con} summarizes this paper.

\section{Preliminaries}\label{sec:pre}
In this section, we introduce RDF and SPARQL patterns,
well-designed patterns, and OPT normal form \cite{P2009Semantics}.

\subsection{RDF} Let ${I}$, ${B}$ and ${L}$ be infinite sets of
\emph{IRIs}, \emph{blank nodes} and \emph{literals}, respectively. These
three sets are pairwise disjoint. We denote the union $I \cup B \cup L$ by
$U$, and elements of $I \cup L$ will be referred to as \emph{constants}.

A triple $(s, p, o) \in ({I}\cup {B}) \times {I} \times ({I} \cup
{B}\cup {L})$ is called an \emph{RDF triple}.
A \emph{basic graph pattern} (BGP) is a set of triple patterns.

\subsection{Semantics of SPARQL patterns}
The semantics of patterns is defined in terms of sets of
so-called \emph{mappings}, which are simply total functions
$\mu \colon S \to U$ on some finite set $S$ of variables.
We denote the domain $S$ of $\mu$ by $\dom \mu$.

Now given a graph $G$ and a pattern $P$, we define the semantics
of $P$ on $G$, denoted by $\semm P G$, as a set of mappings, in
the following manner.
\begin{compactitem}
\item
If $P$ is a triple pattern $(u,v,w)$, then
\begin{center}
$
\semm P G :=
\{\mu \colon \{u,v,w\} \cap V \to U \mid (\mu(u),\mu(v),\mu(w)) \in G\}.
$
\end{center}
Here, for any mapping $\mu$ and any constant $c \in I \cup L$, we
agree that $\mu(c)$ equals $c$ itself. In other
words, mappings are extended to constants according to the
identity mapping.
\item
If $P$ is of the form $P_1 \UNION P_2$, then $\semm P G := \semm {P_1} G \cup \semm {P_2} G$.
\item
If $P$ is of the form $P_1 \ANDAND P_2$, then $\semm P G := \semm {P_1} G \Join \semm {P_2} G$,
where, for any two sets of mappings $\Omega_1$ and $\Omega_2$,
we define
\begin{center}
$
\Omega_1 \Join \Omega = \{\mu_1 \cup \mu_2 \mid \mu_1 \in \Omega_1 \text{ and } \mu_2 \in \Omega_2 \text{ and }\mu_1 \sim \mu_2\}.
$
\end{center}
Here, two mappings $\mu_1$ and $\mu_2$ are called
\emph{compatible}, denoted by $\mu_1 \sim \mu_2$, if
they agree on the intersection of their domains, i.e.,
if for every variable $?x \in \dom {\mu_1} \cap \dom {\mu_2}$, we have
$\mu_1(?x) = \mu_2(?x)$. Note that when $\mu_1$ and $\mu_2$ are
compatible, their union $\mu_1 \cup \mu_2$ is a well-defined
mapping; this property is used in the formal definition above.
\item
If $P$ is of the form $P_1 \OPT P_2$, then
\begin{center}
$
\semm P G := (\semm {P_1} G \Join \semm {P_2} G)
\cup (\semm {P_1} G \smallsetminus \semm {P_2} G),
$
\end{center}
where, for any two sets of mappings $\Omega_1$ and $\Omega_2$,
we define
\begin{center}
$\Omega_1 \smallsetminus \Omega_2$ $ =
\{ \mu_1 \in \Omega_1 \mid \neg \exists \mu_2 \in \Omega_2 :
\mu_1 \sim \mu_2\}$.
\end{center}

\item
If $P$ is of the form $\SELECT_{S}(P_{1})$, then $\semm {P}{G} =
\{\mu|_{S \cap \dom \mu} \mid \mu \in \semm{P_1}G\}$, where $f|_X$
denotes the standard mathematical notion of restriction of a
function $f$ to a subset $X$ of its domain.

\item
Finally, if $P$ is of the form $P_1 \FILTER C$, then $\semm P G := \{\mu \in \semm {P_1} G \mid \mu(C) = \true \}$.

Here, for any mapping $\mu$ and constraint $C$, the evaluation of
$C$ on $\mu$, denoted by $\mu(C)$, is defined in terms of a
three-valued logic with truth values $\true$, $\false$, and $\error$.
Recall that $C$ is a boolean combination of atomic constraints.

For a bound constraint $\bound(?x)$, we define:
\begin{center}
$
\mu(\bound(?x)) =
\begin{cases}
\true & \text{if $?x \in \dom \mu$;} \\
\false & \text{otherwise.}
\end{cases}
$
\end{center}
For an equality constraint $?x=?y$, we define:
\begin{center}
$
\mu(?x=?y) = \begin{cases}
\true & \text{if $?x,?y \in \dom \mu$ and $\mu(?x)=\mu(?y)$;} \\
\false & \text{if $?x,?y \in \dom \mu$ and $\mu(?x)\neq\mu(?y)$;} \\
\error & \text{otherwise.}
\end{cases}
$
\end{center}
Thus, when $?x$ and $?y$ do not both belong to $\dom \mu$, the
equality constraint evaluates to $\error$.
Similarly, for a constant-equality constraint $?x=c$, we define:
\begin{center}
$
\mu(?x=c) = \begin{cases}
\true & \text{if $?x \in \dom \mu$ and $\mu(?x)=c$;} \\
\false & \text{if $?x \in \dom \mu$ and $\mu(?x)\neq c$;} \\
\error & \text{otherwise.}
\end{cases}
$
\end{center}

A boolean combination is then evaluated using the
truth tables given in Table~\ref{truthtable}.
\end{compactitem}
\begin{table}
\caption{Truth tables for the three-valued semantics.}
\label{truthtable}
$$
\begin{array}[t]{cc|cc}
p & q & p \land q & p \lor q \\
\hline
\true & \true & \true & \true\\
\true & \false & \false & \true\\
\true & \error & \error & \true\\
\false & \true & \false & \true\\
\false & \false & \false & \false\\
\false & \error & \false & \error\\
\error & \true & \error & \true\\
\error & \false & \false & \error\\
\error & \error & \error & \error
\end{array}
\qquad
\begin{array}[t]{c|c}
p & \neg p \\
\hline
\true & \false \\
\false & \true \\
\error & \error
\end{array}
$$
\end{table}

\subsection{Well-Designed Pattern}\label{wd}
A $\UNION$-\emph{free} pattern $P$ is \emph{well-designed} if the followings hold:
\begin{itemize}
\item $P$ is safe;
\item for every subpattern $Q$ of form ($Q_{1}$ OPT $Q_{2}$) of $P$ and for every variable $?x$ occurring in $P$, the following condition holds:
\begin{center}
If $?x$ occurs both inside $Q_{2}$ and outside $Q$, then it also occurs in $Q_{1}$.
\end{center}
\end{itemize}

Consider the definition of well-designed patterns, some conceptions can be explained as follows: 
\begin{remark}
In the fragment of and-opt patterns, we exclude $\FILTER$ and $\UNION$ operators and it contains only $\ANDAND$ and $\OPT$ operators at most.
It is obvious that \emph{and-opt} pattern must be $\UNION$-\emph{free} and safe.
\end{remark}

We can conclude that WDAO-patterns are decided by variables in subpattern.
\begin{itemize}
\item \textbf {UNION-\emph{free} Pattern}: $P$ is $\UNION$-free if $P$ is constructed by using only operators $\ANDAND$, $\OPT$, and $\FILTER$.
Every graph pattern $P$ is equivalent to a pattern of the form denoted by ( $P_1 \UNION P_2 \UNION \cdots \UNION P_n$ ).
Each $P_i$ ( $1 \leq i \leq n$ ) is $\UNION$-free.
\item \textbf {Safe} : If the form of ( P FILTER R ) holds the condition of $\var(R) \subseteq \var(P)$, then it is safe.
\end{itemize}

Note that the OPT operator provides really optional left-outer join due to
the weak monotonicity \cite{P2009Semantics}.
A SPARQL pattern P is said to be weakly monotone if for every pair of RDF graphs $G_{1}$, $G_{2}$ such that $G_{1}\subseteq G_{2}$,
it holds that $\semm P {G_1} \sqsubseteq \semm P {G_2}$.
In other words, we assume $\mu_1$ represents $\semm P {G_1}$, and $\mu_2$ represents $\semm P {G_2}$.
Then there exists $\mu'$ such that $\mu_2=\mu_1\cup\mu'$.
Weakly monotone is an important property to characterize the satisfiability of SPARQL \cite{Zhang2015sat}. For instance, consider the pattern $Q_1$ in Section \ref{sec:introduction}, $(?y, \textit{rdf:type}, ?v)$ are really optional.

\subsection{OPT Normal Form}
A UNION-free pattern $P$ is in \emph{OPT normal form} \cite{P2009Semantics}
if $P$ meets one of the following two conditions:
\begin{compactitem}
\item $P$ is constructed by using only the $\ANDAND$ and $\FILTER$
operators;
\item $P = (P_1\ \OPT\ P_2)$ where $P_1$ and $P_2$ patterns are in OPT
normal form.
\end{compactitem}

For instance, the pattern $Q$ aforementioned in Section \ref{sec:introduction} is in OPT normal form.
However, consider the pattern $(((?x, \textit{p}, ?y)\ \OPT\ (?x, \textit{q}, ?z))\ANDAND (?x, \textit{r}, ?z))$ is not in OPT normal form.

\section{Well-Designed And-Opt Tree} \label{sec:wdao}

In this section, we propose the conception of the well-designed and-opt tree (WDAO-tree), any WDAO-pattern can be seen as an WDAO-tree.

\subsection{WDAO-tree Structure}

\begin{definition}[WDAO-tree]\label{well-designed tree}
Let $P$ be a well-designed pattern in OPT normal form. A well-designed tree $T$ based on $P$ is a redesigned parse tree,
which can be defined as follows:
\begin{compactitem}
\item All inner nodes in $T$ are labeled by the $\OPT$ operator and leaves are labeled by BGP.
\item For each subpattern $(P_1\ \OPT \ P_2)$ of $P$, the well-designed
tree $T_1$ of $P_1$ and the well-designed tree $T_2$ of $P_2$ have the same parent node.
\end{compactitem}
\end{definition}

For instance, consider a WDAO-pattern $P$\footnote{We give each OPT operator a
subscript to differentiate them so that readers understand clearly.}
\begin{multline*}
P=(((p_1 \ \ANDAND \ p_3) \ \OPT_2 \ p_2) \ \OPT_1 \ \ \ \ \ \ \ \ \ \ \ \ \ \ \\ ((p_4 \ \OPT_4 \ p_5) \ \OPT_5 (p_6 \ \OPT_6 \ p_7))).
\end{multline*}

\begin{figure}[h]\centering
\begin{tikzpicture}[scale=0.7][nodes={draw}, -]
\node{$\OPT_1$}
child { node {$\OPT_2$}
child { node {$p_1 \ANDAND p_3$}}
child { node {$p_2$} }
}
child [missing]
child [missing]
child { node {$\OPT_3$}
child { node {$\OPT_4$}
child { node {$p_4$}}
child { node {$p_5$}}
}
child [missing]
child { node {$\OPT_5$}
child { node {$p_6$}}
child { node {$p_7$}}
}
};
\end{tikzpicture}
\caption{WDAO-tree\label{fig:wd}}
\end{figure}
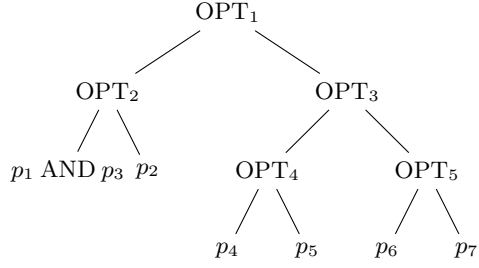

The WDAO-tree $T$ is shown in Figure \ref{fig:wd}.
As shown in this example, BGP - $(p_1 \ANDAND p_3)$ is the exact matching in $P$, which corresponds to the non-optional pattern.
Besides, in WDAO-tree, it is the leftmost leaf in $T$.
We can conclude that the leftmost node in WDAO-tree means the exact matching in well-designed SPARQL query pattern.

\subsection{Rewritting rules over WDAO-tree}
As described in Section \ref{well-designed tree}, WDAO-tree does not contain any OPT operator in its leaves. In this sense, patterns as the form of $Q_1$ in Section \ref{sec:introduction} cannot be transformed into WDAO-tree since it is not OPT normal form.

\begin{proposition}\cite[Theorem 4.11]{P2009Semantics}\label{prop:ONF}
For every UNION-free well-designed pattern $P$, there exists a pattern
$Q$ in OPT normal form such that $P$ and $Q$ are equivalent.
\end{proposition}

In the proof of Proposition \ref{prop:ONF}, we apply three rewriting rules
based on the following equations: let $P, Q, R$ be patterns and $C$ a
constraint,
\begin{compactitem}
\item $(P \OPT R) \ANDAND Q \equiv (P \ANDAND Q) \OPT R$;
\item $P \ANDAND (Q \OPT R) \equiv (P \ANDAND Q) \OPT R$;
\item $(P \OPT R) \FILTER C \equiv (P \FILTER C) \OPT R$.
\end{compactitem}

Intuitively, this lemma states that AND operator can forward and OPT operator can backward in a well-designed pattern with preserving the semantics.
The above three rules can be deployed on a WDAO-tree.
For each WDAO-tree $T$, there exists $T'$ corresponding to $T$ after applying rewriting rules.

\begin{figure}[h]
\center
\subfigure{
\begin{tikzpicture}[scale=0.6][nodes={draw}, -]
\node{$\ANDAND$}
child { node {$P$}
}
child { node {$\OPT$}
child { node {$Q$}}
child { node {$R$} }
};
\end{tikzpicture}}
$\quad \quad \Leftrightarrow \quad \quad$
\subfigure{
\begin{tikzpicture}[scale=0.6][nodes={draw}, -]
\node{$\OPT$}
child { node {$\ANDAND$}
child { node {$P$}}
child { node {$Q$} }
}
child { node {$R$}
};
\end{tikzpicture}}
$\quad \quad \Leftrightarrow \quad \quad$
\subfigure{
\begin{tikzpicture}[scale=0.8][nodes={draw}, -]
\node{$\OPT$}
child { node {$P \ANDAND Q$}
}
child { node {$R$}
};
\end{tikzpicture}}
\caption{rewritting rule-2 \label{rewrite-2}}
\end{figure}
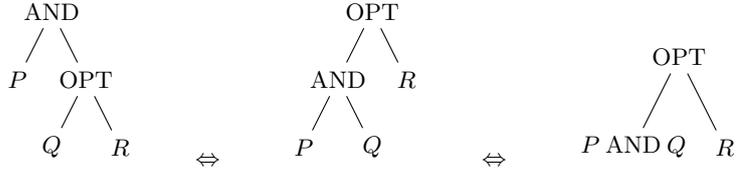

Figure \ref{rewrite-1} and Figure \ref{rewrite-2} have shown that the process of rewriting rules after generating grammar tree and finally WDAO-tree can be obtained. Clearly, WDAO-tree has less height than the grammar tree.

\begin{figure}[h]
\center
\subfigure{
\begin{tikzpicture}[scale=0.6][nodes={draw}, -]
\node{$\ANDAND$}
child { node {$\OPT$}
child { node {$P$}}
child { node {$R$} }
}
child { node {$Q$}
};
\end{tikzpicture}}
$\quad \quad \Leftrightarrow \quad \quad$
\subfigure{
\begin{tikzpicture}[scale=0.6][nodes={draw}, -]
\node{$\OPT$}
child { node {$\ANDAND$}
child { node {$P$}}
child { node {$Q$} }
}
child { node {$R$}
};
\end{tikzpicture}}
$\quad \quad \Leftrightarrow \quad \quad$
\subfigure{
\begin{tikzpicture}[scale=0.8][nodes={draw}, -]
\node{$\OPT$}
child { node {$P \ANDAND Q$}
}
child { node {$R$}
};
\end{tikzpicture}}
\caption{rewritting rule-1 \label{rewrite-1}}
\end{figure}
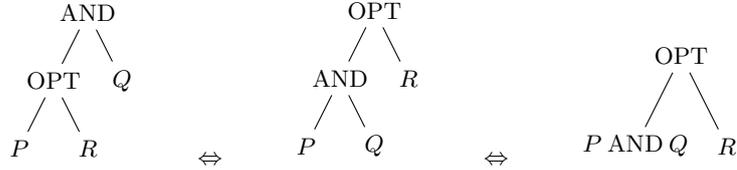

\subsection{WDAO-tree Construction}\label{ctree}

Before constructing WDAO-tree, we recognize query patterns and attachments at first. Then we rewrite query patterns by rewritting rules, which leads to a new pattern. Based on this new pattern, we construct WDAO-tree in the principle of Definition \ref{well-designed tree}.

In the process of the WDAO-tree construction, we firstly build the grammar tree of SPARQL patterns, whose inner node is either $\ANDAND$ operator or $\OPT$ operator. This process is based on recursively putting the left pattern and right pattern of operator in the left node and right node respectively until the pattern does not contain any operator. Then we apply the rewritting rules to the grammar tree in Algorithm \autoref{rewriterules} to build rewriting-tree whose only leaf node is single triple pattern. Different rewritting rules are adopted depending on $\OPT$ operator are $\ANDAND$ operator's left child or right child.
Since WDAO-tree's inner nodes only contain $\ANDAND$ operators,
After getting rewriting-tree, we merge the $\ANDAND$ operators only containing leaf child nodes with its child nodes into new nodes in order to get a WDAO-tree.

The WDAO-tree construction can be executed in PTIME. Given a pattern containing $n$  $\ANDAND$s and $m$ $\OPT$s,
the construction of the grammar tree and rewriting tree have $O(n+m)$ time complexity and $O(nm)$ time complexity, respectively.
Furthermore, the merge of nodes whose parent is $\ANDAND$ has $O(n)$ time complexity.

\begin{center}
\begin{algorithm}[h]
\caption{rewritting rules\label{rewriterules}}
\vspace{.1cm}
\begin{algorithmic}[1]
\Require
GrammarTree with $Root$;
\Ensure
RewriteTree with $Root$;
\State \textbf{while} not all AND.child IS OPT \textbf{do}
\State \textbf{Procedure} ReWriteRules($Root$)
\If{$Root$ IS AND}
\If{$Root$ IS OPT}
\State swap($Root.left$,$Root.right.left$);
\State swap($Root.right$,$Root.right$);
\State swap($Root.left.left$,$Root$);
\State swap($Root.left.right$,$Root.left.right$);
\EndIf
\If{$Root.right$ IS OPT}
\State swap($Root.left$,$Root.left$);
\State swap($Root.right$,$Root.left.left$);
\State swap($Root.left.left$,$Root$);
\State swap($Root.left.right$,$Root.left.right$);
\EndIf
\EndIf
\State \textbf{Procedure} ReWriteRules($Root.left$)
\State \textbf{End Procedure}
\State \textbf{Procedure} ReWriteRules($Root.right$)
\State \textbf{End Procedure}
\State \textbf{End Procedure}
\State \textbf{end while}\\
\Return $Root$;
\end{algorithmic}
\end{algorithm}
\end{center}


\section{PIWD Demonstration} \label{sec:piwd}
In this section, we introduce PIWD, which is a plugin-based framework for well-designed SPARQL.

\subsection{PIWD Overview}
\begin{figure}[h]
\centering
\includegraphics[scale=0.6]{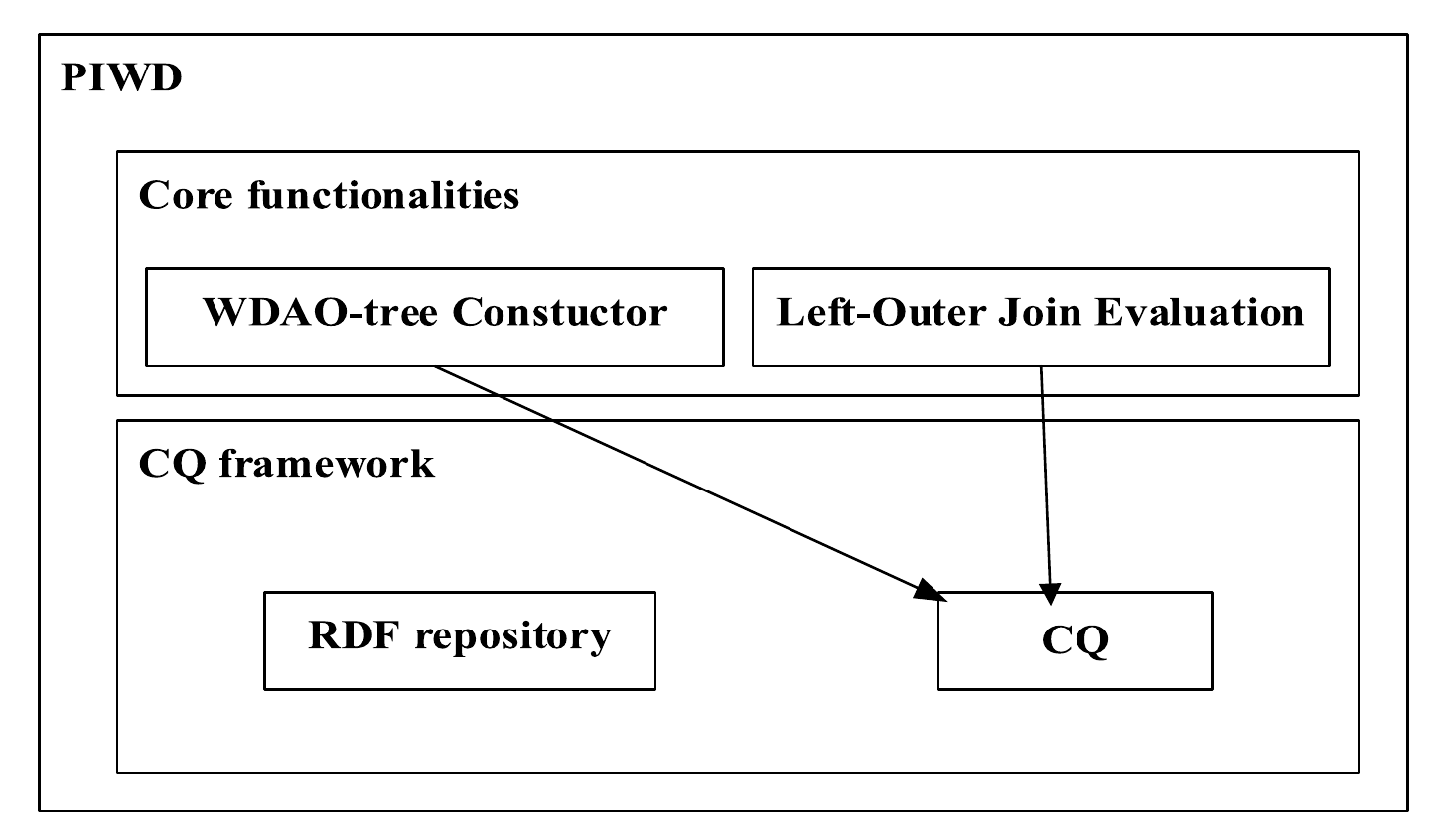}
\caption{PIWD architecture \label{fig:piwd}}
\end{figure}

PIWD is written in Java in a 2-tier design shown in Figure \ref{fig:piwd}.
The bottom layer consists of any BGP query framework which is used as a black box for evaluating BGPs.
Before answering SPARQL queries, the second layer provides the rewriting process and left-outer join evaluation, which lead to the solutions.

BGP query framework supports both query and RDF data management, such as gStore, RDF-3X and so on, which solve the problem of subgraph isomorphism.
PIWD provides the left-outer join between the BGPs.
That is, the problem of answering well-designed SPARQL has been transformed into the problem of subgraph isomorphism and left-outer join between triple patterns.

\subsection{Answering Queries over PIWD}

The query process over PIWD can be described as follows:

Firstly, WDAO-tree is built after rewriting rules on the grammar tree.
Secondly, post-order traversal is applied on WDAO-trees.
The traversal rule is: If the node is a leaf node without the OPT operator, BGP query framework is deployed on it to answer this query and return solutions which is stored in a stack.
If the node is an inner node labeled by the OPT operator, we get the top two elements in the stack and left-outer join them.
We repeat this process until all of WDAO-tree's nodes are visited.
Finally, only one element in the stack is the final solutions.

In the querying processing, BGP query framework serves as a query engine to support queries from leaves in WDAO-trees.
OPT operators take an essential position in the query processing.
Users receive optional solutions based on OPT operators which contribute to the semantic abundance degree since optional solutions are considered in this sense.
In other words, OPT operators lead to the explosive growth of the solution scale.

The query process is described in Algorithm \autoref{queryprocess}.

\begin{center}
\begin{algorithm}[h]
\caption{Query Processing over PIWD \label{queryprocess}}
\begin{algorithmic}[1]
\Require
WDAO-tree with $Root$;
Prefix $prefix$;
$Stack$ to store subresults;
\Ensure
Query result $result$;
\State \textbf{Procedure} TraverseTree($Root$)
\If{$root$ is not null}
\State \textbf{Procedure} TraverseTree($Root \to Lnode$)
\State \textbf{End Procedure}
\State \textbf{Procedure} TraverseTree($Root \to Rnode$)
\State \textbf{End Procedure}
\If{$node$ is not \emph{OPTIONAL}}
\State $subquery$=AssembleQuery($prefix$,$node$);
\State $subresult$=QueryIngStore($subquery$);
\State Push($Stack$\;,\;$subresult$);
\Else
\State $r$=Pop($Stack$);
\State $l$=Pop($Stack$);
\State $result$=$l \; \tiny \textifsym{d|><|} \; r$;
\State Push($Stack$\;,\;$result$);
\EndIf
\EndIf
\State \textbf{End Procedure}
\State $list$=ConvertToList($Stack$);\\
\Return $list$;
\end{algorithmic}
\end{algorithm}
\end{center}

\section{Experiments and Evaluations} \label{sec:experiment}

This section presents our experiments.
The purpose of the experiments is to evaluate the performance
of different WDAO-patterns.
\subsection{Experiments}
\paragraph{Implementations and running environment}
All experiments were carried out on a machine running Linux, which has one CPU with four cores of
2.40GHz, 32GB memory and 500GB disk storage. All of the algorithms were
implemented in Java. gStore\cite{Zou2011gStore,Zou2014gStore} and RDF-3X\cite{Neumann2010The} are used as
the underlying query engines to handle BGPs.
In our experiments, there is no optimization in our OPT operation.

\paragraph{gStore and RDF-3X}
Both gStore and RDF-3X are SPARQL query engines for subgraph matching. gStore stores RDF data in disk-based adjacency lists, whose format is \textit{[vID,vLabel,adjList]}, where \textit{vID} is the vertex ID, \textit{vLabel} is the corresponding URI,
and \textit{adjList} is the list of its outgoing edges and the corresponding neighbor vertices. gStore converts an RDF graph into a data signature graph by encoding each entity and class vertex. Some different
hash functions such as BKDR and AP hash functions are employed to generate signatures, which compose a novel index (called VS$^\ast$-tree).
A filtering rule and efficient search algorithms are developed for subgraph queries over the data signature graph in order to speed up query processing.
gStore can answer exact SPARQL queries and queries with wildcards in a uniform manner. RDF-3X engine is a RISC-style architecture for executing SPARQL queries over large repositories of RDF triple.
Physical design is workload-independent by creating appropriate indexes
over a single giant triples table in RDF-3X. And the query processor is RISC-style by relying mostly on merge joins over sorted index lists.
gStore and RDF-3X have good performances in BGPs since their query methods are based on subgraph matching.

\paragraph{Dataset}
We used LUBM\footnote{http://swat.cse.lehigh.edu/projects/lubm/} as the
dataset in our experiments to investigate the relationship between query response time and dataset scale. LUBM, which features an ontology for the university domain, is a standard benchmark
to evaluate the performance of semantic Web repositories,
In our experiments, we used LUBM1, LUBM50, LUBM100, LUBM150 and
LUBM200 as our query datasets.
The LUBM dataset details in our experiments are shown in Table \ref{lubm}.
\begin{table}
\centering
\caption{LUBM Dataset Details\label{lubm}}
\begin{tabular}{c|r|r}
\hline
Dataset &Number of triples &RDF NT File Size(bytes)\\ \hline 
LUBM1 &103,104 &14,497,954\\ 
LUBM50 &6,890,640 &979,093,554\\ %
LUBM100 &13,879,971 &1,974,277,612\\
LUBM150 &20,659,276 &2,949,441,119\\
LUBM200 &27,643,644 &3,954,351,227\\
\hline
\end{tabular}
\end{table}

\paragraph{SPARQL queries} The queries over LUBM were designed as four different forms, which corresponds to different WDAO-trees.
The details of queries are described in Table \ref{queries}.
Clearly, OPT nesting in $Q_2$ is the most complex among four forms.
Furthermore, we build the $\ANDAND$ operator in each query.

\begin{table}
\centering
\caption{SPARQL queries Details\label{queries}}
\begin{tabular}{c|c|c}
\hline
QueryID &Pattern &OPT amount\\ \hline
$Q_1$ &$(P_1 \ \ANDAND \ P_2 \ \ANDAND P_3)\ \OPT \ P_4$ &1\\
$Q_2$ &$((P_1 \ \ANDAND \ P_2 \ \ANDAND P_3)\ \OPT \ P_4) \ \OPT \ (P_5 \ \OPT \ P_6)$ &3\\
$Q_3$ &$((P_1 \ \ANDAND \ P_2 \ \ANDAND P_3)\ \OPT \ P_4) \ \OPT \ P_5$ &2\\
$Q_4$ &$P_1 \ \OPT \ ((P_2 \ANDAND P_3 \ANDAND P_4) \ \OPT \ P_5)$&2\\
\hline
\end{tabular}
\end{table}

\subsection{Evaluation on PIWD}
\begin{figure}[htbp]
\centering
\subfigure[Performance on gStore] {
\begin{tikzpicture}[scale=0.70]
\begin{axis}[
title={},
xlabel={Dataset scale},
ylabel={Time[ms]},
symbolic x coords={LUBM1,LUBM50,LUBM100,LUBM150,LUBM200},
legend pos=north west,
ymajorgrids=true,
grid style=dashed,
]

\addplot[
color=red,
mark=square*, mark options={fill=black}
]
coordinates {
(LUBM1,1101)(LUBM50,617642)(LUBM100,1329365 )(LUBM150,2126383)(LUBM200,2978237)
};

\addplot[
color=blue,
mark=square*, mark options={fill=red}
]
coordinates {
(LUBM1,1870)(LUBM50,1010965)(LUBM100,2901295)(LUBM150,6623806)(LUBM200,10041836)
};

\addplot[
color=black,
mark=triangle*, mark options={fill=white}
]
coordinates {
(LUBM1,1478)(LUBM50,637128)(LUBM100,1359315)(LUBM150,2191356)(LUBM200,3068692)
};
\addplot[
color=orange,
mark=*, mark options={fill=blue}
]
coordinates {
(LUBM1,1242)(LUBM50,644155)(LUBM100,1456232)(LUBM150,2151811)(LUBM200,3129246)
};
\legend{$Q_1$,$Q_2$,$Q_3$,$Q_4$}

\end{axis}
\end{tikzpicture}}\subfigure[Performance on RDF-3X] {
\begin{tikzpicture}[scale=0.70]
\begin{axis}[
title={},
xlabel={Dataset scale},
ylabel={Time[ms]},
symbolic x coords={LUBM1,LUBM50,LUBM100,LUBM150,LUBM200},
legend pos=north west,
ymajorgrids=true,
grid style=dashed,
]

\addplot[
color=red,
mark=square*, mark options={fill=black}
]
coordinates {
(LUBM1,1231)(LUBM50,625703)(LUBM100,1401782)(LUBM150,2683461)(LUBM200,3496156)
};

\addplot[
color=blue,
mark=square*, mark options={fill=red}
]
coordinates {
(LUBM1,1900)(LUBM50,1245241)(LUBM100,2983394)(LUBM150,7286812)(LUBM200,10852761)
};

\addplot[
color=black,
mark=triangle*, mark options={fill=white}
]
coordinates {
(LUBM1,1499)(LUBM50,640392)(LUBM100,1427392)(LUBM150,2703981)(LUBM200,3672970)
};
\addplot[
color=orange,
mark=*, mark options={fill=blue}
]
coordinates {
(LUBM1,1316)(LUBM50,648825)(LUBM100,1531547)(LUBM150,2791152)(LUBM200,3714042)
};
\legend{$Q_1$,$Q_2$,$Q_3$,$Q_4$}

\end{axis}
\end{tikzpicture}
}
\caption{Query Response Time over LUBM\label{fig:time}}
\end{figure}
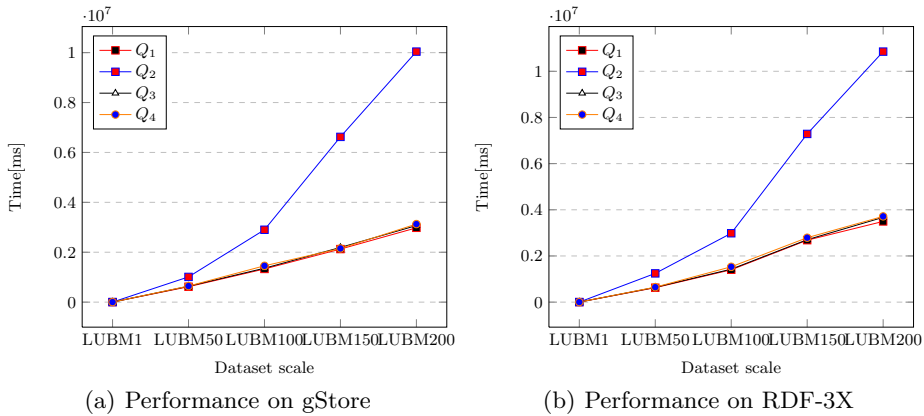

The variation tendencies of query response time are shown in Table \ref{tab:time-1}, Table \ref{tab:time-2} and Figure \ref{fig:time}.
Query efficiency is decreased with higher
response time when OPT nesting becomes more complex.
Furthermore, there has been a significant increase in query response time when the dataset scale grows up.
For instance, we observe $Q_2$, which corresponds to the most complex pattern in our four experimental SPARQL patterns.
When the dataset is ranging from LUBM100 to LUBM200, its query response time extends more than five times even though the dataset scale extends two times. In this sense, OPT nesting complexity in WDAO-patterns influences query response time especially for large dataset scale.

\begin{table}
\centering
\caption{Query Response Time[ms] on gStore\label{tab:time-1}}
\begin{tabular}{c|r|r|r|r|r}
\hline
&LUBM1 &LUBM50 &LUBM100 &LUBM150 &LUBM200\\ \hline
$Q_1$ &1,101 &617,642 &1,329,365 &2,126,383 &2,978,237 \\
$Q_2$ &1,870 &1,010,965 &2,901,295 &6,623,806 &10,041,836 \\
$Q_3$ &1,478 &637,128 &1,359,315 &2,191,356 &3,068,692 \\
$Q_4$ &1,242 &644,155 &1,456,232 &2,151,811 &3,129,246\\
\hline
\end{tabular}
\end{table}

\begin{table}
\centering
\caption{Query Response Time[ms] on RDF-3X\label{tab:time-2}}
\begin{tabular}{c|r|r|r|r|r}
\hline
&LUBM1 &LUBM50 &LUBM100 &LUBM150 &LUBM200\\ \hline
$Q_1$ &1,231 &625,703 &1,401,782 &2,683,461 &3,496,156 \\
$Q_2$ &1,900 &1,245,241 &2,983,394 &7,286,812 &10,852,761 \\
$Q_3$ &1,499 &640,392 &1,427,392 &2,703,981 &3,672,970 \\
$Q_4$ &1,316 &648,825 &1,531,547 &2,791,152 &3,714,042\\
\hline
\end{tabular}
\end{table}

\section{Related works}\label{sec:related}
In this section, we survey related works in the following three areas: BGP query evaluation algorithms, well-designed SPARQL and BGP query evaluation frameworks.

BGP query algorithms have been developed for many years. Existing algorithms mainly focus on finding all embedding in a single large graph, such as ULLmann\cite{Ullmann1976An}, VF2\cite{Luigi2004A}, QUICKSI\cite{Shang2008Taming}, GraphQL\cite{He2010Query}, SPath\cite{Zhao2010On}, STW\cite{Hongzhi2012Efficient} and TurboIso\cite{Han2013Turbo}. Some optimization method has been adapted in these techniques, such as adjusting matching order, pruning out the candidate vertices. However, the evaluation of well-designed SPARQL is not equivalent to the BGP query evaluation problem since there exists inexact matching.

It has been shown that the complexity of the evaluation problem for the well-designed fragment is coNP-complete\cite{P2009Semantics}.
The quasi well-designed pattern trees (QWDPTs), which are undirected and ordered, has been proposed \cite{Letelier2012Static}.
This work aims at the analysis of containment and equivalence of well-designed pattern. Efficient evaluation and semantic optimization of WDPT have been proposed in \cite{barcelo2015efficient}. Sparm is a tool for SPARQL analysis and manipulation in \cite{letelier2012spam}.
Above-mentioned all aim at checking well-designed patterns or complexity analysis without evaluation on well-designed patterns.
Our WDAO-tree is different from QWDPTs in structure and it emphasizes reconstructing query plans. The OPT operation optimization has been proposed in \cite{Atre2015Left}, which is different from our work since
our work aims to handle a plugin in any BGP query engine in order to deal with WDAO-patterns in SPARQL queries.

RDF-3X\cite{Neumann2010The}, TripleBit\cite{Yuan2013TripleBit}, SW-Store\cite{abadi2009sw}, Hexastore\cite{Weiss2008Hexastore} and gStore\cite{Zou2011gStore,Zou2014gStore} have high performance in BGPs.
RDF-3X create indexes in the form of B+ tree, as well as TripleBit in the form of ID-Chunk. All of them have efficient performance since they concentrate on the design of indexing or storage. However, they can only support exact SPARQL queries, since they replace all literals
(in RDF triples) by ids using a mapping dictionary.
In other words, they cannot support WDAO-patterns well.
Virtuoso\cite{Erling2009RDF} and MonetDB\cite{Boncz2005MonetDB} support open-source and commercial services.
Jena\cite{Carroll2004Jena} and Sesame\cite{Broekstra2002Sesame} are free open source Java frameworks for building semantic web and Linked Data applications, which focus on SPARQL parse without supporting large-scale date. Our work is independent on these BGP query frameworks, and any BGP query engine is adaptable for our plugin.

\section{Conclusion}\label{sec:con}
In this paper, we have presented PIWD, which is a plugin adaptable for any BGP query framework to handle WDAO-patterns. Theoretically, PIWD rebuilds the query evaluation plan based on WDAO-trees. After employing BGP query framework on WDAO-trees, PIWD supports the left-outer join operation between triple patterns. Our experiments show that PIWD can deal with complex and multi-level nested WDAO-patterns. In the future, we will further handle other non-well-designed patterns and deal with more operations such as UNION. Besides, we will consider OPT operation optimization to improve efficiency of PIWD and implement our framework on distributed {RDF} graphs by applying the distributed gStore \cite{DBLP:journals/vldb/PengZO0Z16}.

\section*{Acknowledgments}
This work is supported by the programs of the National Key Research and Development Program of China (2016YFB1000603), the National Natural Science Foundation of China (NSFC) (61502336), and the open funding project of Key Laboratory of Computer Network and Information Integration (Southeast University), Ministry of Education (K93-9-2016-05). Xiaowang Zhang is supported by Tianjin Thousand Young Talents Program and the project-sponsored by School of Computer
Science and Technology in Tianjin University.

\end{document}